\documentclass[prd,aps,floatfix,nofootinbib,11 pt]{revtex4}
\usepackage{amsmath,graphicx,color,epsfig}

\input{tcilatex}

\begin{document}

\title{Communication aspects of a three-player Prisoner's Dilemma quantum
game}
\author{M. Ramzan\thanks{%
mramzan@phys.qau.edu.pk} and M. K. Khan}
\address{Department of Physics Quaid-i-Azam University \\
Islamabad 45320, Pakistan}

\begin{abstract}
We present a quantization scheme for a three-player Prisoner's Dilemma game.
It is shown that entanglement plays a dominant role in the three-player
quantum game. Four different types of payoffs are identified on the basis of
different combinations of initial state and measurement basis entanglement
parameters. A relation among these different payoffs is also established. We
also study the communication aspects of the three-player game. By exploiting
different combinations of initial state and measurement basis entanglement
parameters, we establish a relationship for the information shared among the
parties. It is seen that the strategies of the players act as carriers of
information in quantum games.
\end{abstract}

\maketitle

\address{Department of Physics Quaid-i-Azam University \\
Islamabad 45320, Pakistan}

\address{Department of Physics Quaid-i-Azam University \\
Islamabad 45320, Pakistan}


\vspace*{1.0cm}

\vspace*{1.0cm}

%\date{\today}

%\newpage

\section{Introduction}

Recent development in quantum computation and quantum information theory
\cite{Nelson,Dirk} prompted the scope of game theory to extend it to the
quantum world. Meyer \cite{Meyer} discussed a connection between quantum
games and quantum information processing. Most of the research on quantum
games has lacked a direct connection to quantum information processing.
Quantum game theory has been extensively studied by a number of authors in
the last few years [4-6]. The role of the initial quantum state entanglement
is an interesting feature in quantum games. However, the importance of the
payoff operators used by the arbiter to perform measurement is also
important as addressed in ref. \cite{Nawaz}. The authors have investigated
the role of measurement basis in quantum games by taking the two-player
Prisoner's Dilemma game as an example. Lee et al. \cite{Lee2}, have studied
the problem of quantum state estimation and quantum cloning using a game
theoretic perspective.

The Prisoner's Dilemma is a widely known example in classical game theory.
The study of multi-player quantum games could be of great importance from
both a theoretical and a practical point of view and can exhibit interesting
results in comparison to two player games. A model of two player Prisoner's
Dilemma quantum game was developed by Eisert \cite{Eisert} in which the
paradox in the classical Prisoner's Dilemma was solved in a maximally
entangled state. Quantum Prisoner's Dilemma has been experimentally
demonstrated by using a nuclear magnetic resonance (NMR) quantum computer
\cite{Du1}. Recently, Prevedel et al. have experimentally demonstrated the
application of a measurement-based protocol to realize a quantum version of
the Prisoner's Dilemma based on entangled photonic cluster states and
constituted the first realization of a quantum game in the context of one
way quantum computing \cite{Prev}. The investigations of multi-player and
multi-choice quantum games, [12-14] and continuous-variable quantum games
\cite{Li} have also been pursued in the recent years.

With recent interest in quantum computing and quantum information theory, we
explore that quantum game theory may be useful in studying the quantum
communication, since it can be considered as a game where the objective is
to maximize the effective communication. Motivated from our recent paper on
two player quantum games \cite{Ramzan}, we extend our work here to the case
of a three-player Prisoner's Dilemma quantum game with measuring basis taken
as entangled. Motivation of three-player Prisoner's Dilemma quantum game is
that more information can be carried by each party that may increase the
communication of information. Further more, this work may provide a better
insight in the study of quantum games from the quantum information and
quantum communication perspective. Based on the work discussed in ref. \cite%
{Kawakami,Orlin}, we have attempted to relate the quantum game theory with
quantum information theory by investigating the communication aspects of a
three-player Prisoner's Dilemma quantum game. Kawakami \cite{Kawakami} has
studied the communication and information carriers in quantum games. He has
shown that communications in quantum games can be used to solve problems
that cannot be solved by using communications in classical games.

In this paper, we present a quantization scheme for three-player Prisoner's
Dilemma game using entangled measuring basis. We study the communication
aspects of the three-player Prisoner's Dilemma game by using the players
returns. Based on the flow of information (communication) between players,
as evident from the payoff matrix, we establish a relationship for
information shared among the parties, for different combinations of initial
state and measurement basis entanglement parameters $\gamma \in \lbrack
0,\pi /2]$ and $\delta \in \lbrack 0,\pi /2]$ respectively. Here, $\delta =0$
means that the measurement basis are unentangled i.e. in product form and $%
\delta =\pi /2$ means that it is maximally entangled. Similarly, $\gamma =0$
means that the game is initially unentangled and $\gamma =\pi /2$ means that
it is maximally entangled. We show that the strategies of the players and
their payoffs act as information carriers between the players. We establish
a relationship among different payoffs on the basis of different
combinations of initial state and measurement basis entanglement parameters $%
\delta $ and $\gamma $ respectively, as studied in ref. \cite{Nawaz}. The
relation among different quantum payoffs is similar to the relation among
classical capacities of the quantum channels \cite{King}. In addition, we
also establish a relationship among the information shared between the
parties for different combinations of initial state and measurement basis
entanglement parameters.

\section{Three-player quantization scheme}

The three-player Prisoner's Dilemma game is similar to the two-player
situation. In a three-player Prisoner's Dilemma game, the players are
arrested under the suspicion of committing a crime, say, robbing a bank.
Similar to the two player game, they are interrogated in separate cells
without communicating with each other. The two possible moves for each
prisoner are, to cooperate $(C)$ or to defect $(D).$ The payoff table for
the three-player Prisoner's Dilemma game is shown in table 1 \cite{Du2}. The
game is symmetric for the three players, and the strategy ($D)$ dominates
the strategy ($C)$ for all the three players. Since the selfish players
prefer to choose ($D)$ as optimal strategy, the unique Nash equilibrium is ($%
D,D,D$) with payoffs ($1,1,1$). This is a Pareto inferior outcome, since ($%
C,C,C$) with payoffs ($3,3,3$) would be better for all three players. This
situation is the very catch of the dilemma and is the same as the two-player
version of this game.

In our scheme, Alice, Bob and a third player, Charlie, join the game. In
this game, an arbiter prepares an initial quantum state and passes it on to
the players. After applying their strategies, the players return the state
to the arbiter who then announces the payoffs by performing a measurement.
Let us suppose that the initial quantum state shared between the three
prisoners, consistent with \cite{Ramzan,Nawaz2}, is of the form%
\begin{equation}
\left\vert \psi _{in}\right\rangle =\cos \frac{\gamma }{2}\left\vert
000\right\rangle +i\sin \frac{\gamma }{2}\left\vert 111\right\rangle ,
\label{initial}
\end{equation}%
where $0\leq \gamma \leq \pi /2$ corresponds to the entanglement of the
initial state. Here in this case players can locally manipulate their
individual qubits. The possible outcomes of the classical strategies ($C$)\
and ($D$)\ are assigned the two basis vector $\left\vert 0\right\rangle $
and $\left\vert 1\right\rangle $ in the Hilbert space. The strategies of the
players can be represented by the unitary operator $U_{k}$ as defined in
ref. \cite{Ramzan}%
\begin{equation}
U_{k}=\cos \frac{\theta _{k}}{2}R_{k}+\sin \frac{\theta _{k}}{2}P_{k},
\label{unit}
\end{equation}%
where $k=A$, $B$ \& $C$\ correspond to Alice, Bob and Charlie respectively
and $R_{k}$, $P_{k}$\emph{\ }are the unitary operators defined as

\begin{align}
R_{A}\left\vert 0\right\rangle & =e^{i\alpha _{A}}\left\vert 0\right\rangle
\text{\qquad \qquad \qquad }R_{A}\left\vert 1\right\rangle =e^{-i\alpha
_{A}}\left\vert 1\right\rangle  \notag \\
P_{A}\left\vert 0\right\rangle & =e^{i\left( \frac{\pi }{2}-\beta
_{A}\right) }\left\vert 1\right\rangle \text{\qquad \qquad }P_{A}\left\vert
1\right\rangle =e^{i\left( \frac{\pi }{2}+\beta _{A}\right) }\left\vert
0\right\rangle  \notag \\
R_{B}\left\vert 0\right\rangle & =e^{i\alpha _{B}}\left\vert 0\right\rangle
\text{\qquad \qquad \qquad }R_{B}\left\vert 1\right\rangle =e^{-i\alpha
_{B}}\left\vert 1\right\rangle  \notag \\
P_{B}\left\vert 0\right\rangle & =e^{i\left( \frac{\pi }{2}-\beta
_{B}\right) }\left\vert 1\right\rangle \text{\qquad \qquad }P_{B}\left\vert
1\right\rangle =e^{i\left( \frac{\pi }{2}+\beta _{B}\right) }\left\vert
0\right\rangle  \notag \\
R_{C}\left\vert 0\right\rangle & =e^{i\alpha _{C}}\left\vert 0\right\rangle
\text{\qquad \qquad \qquad }R_{C}\left\vert 1\right\rangle =e^{-i\alpha
_{C}}\left\vert 1\right\rangle  \notag \\
P_{C}\left\vert 0\right\rangle & =e^{i\left( \frac{\pi }{2}-\beta
_{C}\right) }\left\vert 1\right\rangle \text{\qquad \qquad }P_{C}\left\vert
1\right\rangle =e^{i\left( \frac{\pi }{2}+\beta _{C}\right) }\left\vert
0\right\rangle ,  \label{unitdef}
\end{align}%
where $0\leq \theta _{k}\leq \pi ,-\pi \leq \{\alpha _{k},\beta _{k}\}\leq
\pi .$ By application of the local operators of the players, the initial
state given in equation (\ref{initial}) transforms to
\begin{equation}
\rho _{f}=(U_{A}\otimes U_{B}\otimes U_{C})\rho _{in}(U_{A}\otimes
U_{B}\otimes U_{C})^{\dagger },  \label{final}
\end{equation}%
where $\rho _{in}=\left\vert \psi _{in}\right\rangle \left\langle \psi
_{in}\right\vert $ is the initial density matrix for the quantum state. The
operators used by the arbiter to determine the payoffs for Alice, Bob and
Charlie are
\begin{eqnarray}
P^{k}
&=&\$_{000}^{k}P_{000}+\$_{001}^{k}P_{001}+\$_{110}^{k}P_{110}+%
\$_{010}^{k}P_{010}  \notag \\
&&+\$_{101}^{k}P_{101}+\$_{011}^{k}P_{011}+\$_{100}^{k}P_{100}+%
\$_{111}^{k}P_{111}
\end{eqnarray}%
where
\begin{eqnarray}
P_{000} &=&\left\vert \psi _{000}\right\rangle \left\langle \psi
_{000}\right\vert ,\qquad \left\vert \psi _{000}\right\rangle =\cos \frac{%
\delta }{2}\left\vert 000\right\rangle +i\sin \frac{\delta }{2}\left\vert
111\right\rangle  \notag \\
P_{111} &=&\left\vert \psi _{111}\right\rangle \left\langle \psi
_{111}\right\vert ,\qquad \left\vert \psi _{111}\right\rangle =\cos \frac{%
\delta }{2}\left\vert 111\right\rangle +i\sin \frac{\delta }{2}\left\vert
000\right\rangle  \notag \\
P_{001} &=&\left\vert \psi _{001}\right\rangle \left\langle \psi
_{001}\right\vert ,\qquad \left\vert \psi _{001}\right\rangle =\cos \frac{%
\delta }{2}\left\vert 001\right\rangle +i\sin \frac{\delta }{2}\left\vert
110\right\rangle  \notag \\
P_{110} &=&\left\vert \psi _{110}\right\rangle \left\langle \psi
_{110}\right\vert ,\qquad \left\vert \psi _{110}\right\rangle =\cos \frac{%
\delta }{2}\left\vert 110\right\rangle +i\sin \frac{\delta }{2}\left\vert
001\right\rangle  \notag \\
P_{010} &=&\left\vert \psi _{010}\right\rangle \left\langle \psi
_{010}\right\vert ,\qquad \left\vert \psi _{010}\right\rangle =\cos \frac{%
\delta }{2}\left\vert 010\right\rangle -i\sin \frac{\delta }{2}\left\vert
101\right\rangle  \notag \\
P_{101} &=&\left\vert \psi _{101}\right\rangle \left\langle \psi
_{101}\right\vert ,\qquad \left\vert \psi _{101}\right\rangle =\cos \frac{%
\delta }{2}\left\vert 101\right\rangle -i\sin \frac{\delta }{2}\left\vert
010\right\rangle  \notag \\
P_{011} &=&\left\vert \psi _{011}\right\rangle \left\langle \psi
_{011}\right\vert ,\qquad \left\vert \psi _{011}\right\rangle =\cos \frac{%
\delta }{2}\left\vert 011\right\rangle -i\sin \frac{\delta }{2}\left\vert
100\right\rangle  \notag \\
P_{100} &=&\left\vert \psi _{100}\right\rangle \left\langle \psi
_{100}\right\vert ,\qquad \left\vert \psi _{100}\right\rangle =\cos \frac{%
\delta }{2}\left\vert 100\right\rangle -i\sin \frac{\delta }{2}\left\vert
011\right\rangle  \label{mbasis}
\end{eqnarray}%
where $0\leq \delta \leq \pi /2$ and $\$_{lmn}^{k}$ are the elements of
payoff matrix as given in table 1. Since quantum mechanics is a
fundamentally probabilistic theory, the strategic notion of the payoff is
the expected payoff. The players after their actions, that leave the game in
a state given in equation (\ref{final}), forward their qubits to the arbiter
for the final projective measurement, for example, in the computational
basis as given in equation (\ref{mbasis}), who determines their payoffs (as
shown in figure 1). The payoffs for the players can be obtained as the mean
values of the payoff operators%
\begin{equation}
\$^{k}(\theta _{k},\alpha _{A},\beta _{A})=\text{Tr}(P^{k}\rho _{f}),
\end{equation}%
where Tr represents the trace of the matrix. Using equations (1) to (7), the
payoffs of the three players are given by%
\begin{eqnarray}
&&\left. \$^{k}(\theta _{k},\alpha _{k},\beta _{k})=\right.  \notag \\
&&c_{A}c_{B}c_{C}[\eta _{1}\$_{000}^{k}+\eta
_{2}\$_{111}^{k}+(\$_{000}^{k}-\$_{111}^{k})\xi \cos 2(\alpha _{A}+\alpha
_{B}+\alpha _{C})]  \notag \\
&&+s_{A}s_{B}s_{C}[\eta _{2}\$_{000}^{k}+\eta
_{1}\$_{111}^{k}-(\$_{000}^{k}-\$_{111}^{k})\xi \cos 2(\beta _{A}+\beta
_{B}+\beta _{C})]  \notag \\
&&+c_{A}c_{B}s_{C}[\eta _{1}\$_{001}^{k}+\eta
_{2}\$_{110}^{k}+(\$_{001}^{k}-\$_{110}^{k})\xi \cos 2(\alpha _{A}+\alpha
_{B}-\beta _{C})]  \notag \\
&&+s_{A}s_{B}c_{C}[\eta _{2}\$_{001}^{k}+\eta
_{1}\$_{110}^{k}-(\$_{001}^{k}-\$_{110}^{k})\xi \cos 2(\beta _{A}+\beta
_{B}-\alpha _{C})]  \notag \\
&&+s_{A}c_{B}c_{C}[\eta _{1}\$_{100}^{k}+\eta
_{2}\$_{011}^{k}+(\$_{100}^{k}-\$_{011}^{k})\xi \cos 2(\alpha _{B}+\alpha
_{C}-\beta _{A})]  \notag \\
&&+c_{A}s_{B}s_{C}[\eta _{2}\$_{100}^{k}+\eta
_{1}\$_{011}^{k}-(\$_{100}^{k}-\$_{011}^{k})\xi \cos 2(\beta _{B}+\beta
_{C}-\alpha _{A})]  \notag \\
&&+s_{A}c_{B}s_{C}[\eta _{1}\$_{101}^{k}+\eta
_{2}\$_{010}^{k}+(\$_{101}^{k}-\$_{010}^{k})\xi \cos 2(\beta _{A}+\beta
_{C}-\alpha _{B})]  \notag \\
&&+c_{A}s_{B}c_{C}[\eta _{2}\$_{101}^{k}+\eta
_{1}\$_{010}^{k}-(\$_{101}^{k}-\$_{010}^{k})\xi \cos 2(\alpha _{A}+\alpha
_{C}-\beta _{B})]  \notag \\
&&+\frac{1}{8}(\cos ^{2}(\delta /2)-\sin ^{2}(\delta
/2))[\$_{000}^{k}-\$_{111}^{k}-\$_{001}^{k}+\$_{110}^{k}-\$_{010}^{k}+%
\$_{101}^{k}+\$_{011}^{k}-\$_{100}^{k}]\times  \notag \\
&&\sin (\gamma )\sin (\theta _{1})\sin (\theta _{2})\sin (\theta _{3})\cos
(\alpha _{A}+\alpha _{B}+\alpha _{C}-\beta _{A}-\beta _{B}-\beta _{C})
\notag \\
&&+[[\$_{000}^{k}-\$_{111}^{k}]\sin (\delta )\sin (\theta _{1})\sin (\theta
_{2})\sin (\theta _{2})\cos (\alpha _{A}+\alpha _{B}+\alpha _{C}-\beta
_{A}-\beta _{B}-\beta _{C})  \notag \\
&&+[\$_{110}^{k}-\$_{001}^{k}]\sin (\delta )\sin (\theta _{1})\sin (\theta
_{2})\sin (\theta _{2})\cos (\alpha _{A}+\alpha _{B}-\alpha _{C}+\beta
_{A}+\beta _{B}-\beta _{C})  \notag \\
&&+[\$_{010}^{k}-\$_{101}^{k}]\sin (\delta )\sin (\theta _{1})\sin (\theta
_{2})\sin (\theta _{2})\cos (\alpha _{A}-\alpha _{B}+\alpha _{C}+\beta
_{A}-\beta _{B}+\beta _{C})  \notag \\
&&+[\$_{100}^{k}-\$_{011}^{k}]\sin (\delta )\sin (\theta _{1})\sin (\theta
_{2})\sin (\theta _{2})\cos (\alpha _{A}-\alpha _{B}-\alpha _{C}+\beta
_{A}-\beta _{B}-\beta _{C})]\times  \notag \\
&&[\frac{1}{8}(\cos ^{2}(\gamma /2)-\sin ^{2}(\gamma /2))]  \label{payoff}
\end{eqnarray}%
where
\begin{eqnarray}
\eta _{1} &=&\cos ^{2}(\gamma /2)\cos ^{2}(\delta /2)+\sin ^{2}(\gamma
/2)\sin ^{2}(\delta /2)  \notag \\
\eta _{2} &=&\sin ^{2}(\gamma /2)\cos ^{2}(\delta /2)+\sin ^{2}(\delta
/2)\cos ^{2}(\gamma /2)  \notag \\
\xi &=&\frac{1}{2}\sin (\delta )\sin (\gamma )  \notag \\
c_{k} &=&\cos ^{2}\frac{\theta _{k}}{2}  \notag \\
s_{k} &=&\sin ^{2}\frac{\theta _{k}}{2}
\end{eqnarray}%
The payoffs for the three players can be found by substituting the
appropriate values for $\$_{lmn}^{k}$ into equation (\ref{payoff}). The
elements of classical payoff matrix for the Prisoner's Dilemma game are
given in table 1. Our results are consistent with ref. \cite{Du2} and can be
easily checked from equation (\ref{payoff}), when all the three players
resort to their Nash equilibrium strategies.

\section{Communication Scenario}

Let us start with an analysis of the communication aspects of the quantized
Prisoner's Dilemma game. The communication aspect of quantum games is
similar to the dense coding \cite{Coding}, in the sense that, we can
transmit two bits of classical information by sending only one qubit with
the help of entanglement while the sender and the receiver share an
entangled quantum state. Motivation of three-player quantum game is that
more information can be carried by each party that may increase the
information flux in comparison to the standard two-player version of the
Prisoner's Dilemma game. Further more, the realization of the communication
is due to the advantage of quantum strategies and quantum entanglement. In
our approach, each prisoner has his/her private qubit and applies the
unitary transformation to this. Their arbiter gives a payoff to each of them
based on a measured result of each qubit. Unitary transformations are
strategies for prisoners which play a key role in constructing the payoff
matrix. Here we consider that the strategies of prisoners are represented by
the local operators of Alice, Bob and Charlie as given in equation (\ref%
{unit}). Let Alice, Bob and Charlie agree on that Alice performs the
following four unitary operations out of the set $U_{A}\left( \theta
_{A},\alpha _{A},\beta _{A}\right) ,$\ as given in the below equation, on
her qubit%
\begin{eqnarray}
U_{A}\left( 0,0,0\right) &\Rightarrow &00  \notag \\
U_{A}\left( \frac{\pi }{3},\frac{\pi }{2},\frac{\pi }{2}\right) &\Rightarrow
&01  \notag \\
U_{A}\left( \frac{\pi }{2},\frac{\pi }{2},\frac{\pi }{2}\right) &\Rightarrow
&10  \notag \\
U_{A}\left( \pi ,\pi ,\pi \right) &\Rightarrow &11  \label{unitary}
\end{eqnarray}%
where $00$, $01$, $10$, $11$ represent the exchange of two bits of
information. In order to obtain the classical payoff matrix, we consider the
case of a restricted game, where Alice is allowed to get benefit from the
quantum phases whereas Bob and Charlie are restricted to do so with fixed
phase change by setting $\alpha _{B}=\alpha _{C}=0$ and $\beta _{B}=\beta
_{C}=\pi /2$. Thus, restricting Bob and Charlie to only apply $\theta
_{B(C)}=0$ or $\pi $ as their set of strategies, utilizing which one can
construct the classical payoff matrix for the three-player Prisoner's
Dilemma game as given in table 1. As a result of measurement, Bob and
Charlie can extract the information about the strategy applied by Alice from
their payoffs by mutual understanding that they will apply the same strategy
i.e. either $\theta _{B(C)}=0$ or $\theta _{B(C)}=\pi $, such a cooperation
between the users can avoid corruption in quantum communication. Because,
application of the unitary operators changes not only the value of a qubit,
but also its phase (amplitude). That results a communication of two bits of
information by two local one-qubit operations among the parties (as seen
from table 2).

For example, let Bob and Charlie apply $\theta _{B(C)}=0$, and gain the
payoffs $2,2$ respectively and they can easily find that the decision of
Alice was $U_{A}\left( \pi ,\pi ,\pi \right) $ with payoff $5$ as can be
seen from table $2$. In this case, information which is exchanged between
them through the arbiter is represented as $2$ bits, to determine one of the
four possibilities.

\section{Relationship between payoffs and information}

Quantum payoffs can be divided into four different categories on the basis
of four different combinations of the initial state and measurement basis
entanglement parameters $\gamma $ and $\delta $. These different situations
arise due to the possibility of having a product or entangled initial state
and then applying a product or entangled basis for the measurement \cite%
{Patil,Kim}. Here, we will use the subscripts $E$ and $P$ which correspond
to the entangled and product basis being used for quantum payoffs
respectively. The four different types of payoffs can be categorized as%
\newline
\textbf{Case (a)} When $\delta =\gamma =0$ (i.e. initial quantum state used
is in the product form, and product basis are used for measurement to
determine the payoffs), the game becomes classical and each player plays the
strategy $C$, with probability $\cos ^{2}(\theta _{k}/2)$ and the payoffs
for the players at the Nash equilibrium become%
\begin{equation}
\$_{PP}^{k}(\theta _{k}=\pi )=1
\end{equation}%
\textbf{Case (b)} When $\gamma =0,$ $\delta \neq 0$ (i.e. initial quantum
state used is in the product form, and entangled basis are used for
measurement to determine the payoffs)\ the players' payoff remains less than
3 at\ the two Nash equilibria arising at $\theta _{k}=0$ and $\pi /2$ which
reads%
\begin{eqnarray}
\$_{PE}^{k}(\theta _{k} &=&\pi /2,\alpha _{A}=\pi ,\beta _{A}=\pi )<3 \\
\$_{PE}^{k}(\theta _{k} &=&0,\alpha _{A}=\pi ,\beta _{A}=\pi )<3
\end{eqnarray}%
\textbf{Case (c) }When $\delta =0,$ $\gamma \neq 0$ (i.e. initial quantum
state is entangled, and product basis are used for measurement to determine
the payoffs),\ the players' payoff again remains less than 3 at\ the two
Nash equilibria and is given as%
\begin{eqnarray}
\$_{EP}^{k}(\theta _{k} &=&\pi /2,\alpha _{A}=\pi ,\beta _{A}=\pi )<3 \\
\$_{EP}^{k}(\theta _{k} &=&0,\alpha _{A}=\pi ,\beta _{A}=\pi )<3
\end{eqnarray}%
\textbf{Case (d)} When $\gamma =\delta =\pi /2$ (i.e. initial quantum state
is in entangled form and entangled basis are used for measurement to
determine the payoffs), the players' payoff can be obtained from%
\begin{eqnarray}
\$_{EE}^{k}(\theta _{k},\alpha _{A},\beta _{A}) &=&  \notag \\
&&\frac{c_{A}c_{B}c_{C}}{2}[(\$_{000}^{k}+\$_{111}^{k})+(\$_{000}^{k}-%
\$_{111}^{k})\xi \cos 2(\alpha _{A})]  \notag \\
&&+\frac{s_{A}s_{B}s_{C}}{2}[(\$_{000}^{k}+\$_{111}^{k})-(\$_{000}^{k}-%
\$_{111}^{k})\xi \cos 2(\beta _{A})]  \notag \\
&&+\frac{c_{A}c_{B}s_{C}}{2}[(\$_{001}^{k}+\$_{110}^{k})+(\$_{001}^{k}-%
\$_{110}^{k})\xi \cos 2(\alpha _{A})]  \notag \\
&&+\frac{s_{A}s_{B}c_{C}}{2}[(\$_{001}^{k}+\$_{110}^{k})-(\$_{001}^{k}-%
\$_{110}^{k})\xi \cos 2(\beta _{A})]  \notag \\
&&+\frac{s_{A}c_{B}c_{C}}{2}[(\$_{100}^{k}+\$_{011}^{k})+(\$_{100}^{k}-%
\$_{011}^{k})\xi \cos 2(\beta _{A})]  \notag \\
&&+\frac{c_{A}s_{B}s_{C}}{2}[(\$_{100}^{k}+\$_{011}^{k})-(\$_{100}^{k}-%
\$_{011}^{k})\xi \cos 2(\alpha _{A})]  \notag \\
&&+\frac{s_{A}c_{B}s_{C}}{2}[(\$_{101}^{k}+\$_{010}^{k})+(\$_{101}^{k}-%
\$_{010}^{k})\xi \cos 2(\beta _{A})]  \notag \\
&&+\frac{c_{A}s_{B}c_{C}}{2}[(\$_{101}^{k}+\$_{010}^{k})-(\$_{101}^{k}-%
\$_{010}^{k})\xi \cos 2(\alpha _{A})]
\end{eqnarray}%
The payoffs, when the three players play their Nash equilibrium strategies
become%
\begin{equation}
\$_{EE}^{k}(\theta _{k}=0,\alpha _{A}=\pi ,\beta _{A}=\pi )=3
\end{equation}%
From the above four cases one can establish the following relation among the
four payoff values as
\begin{equation}
\$_{PP}^{k}<\$_{PE}^{k}=\$_{EP}^{k}<\$_{EE}^{k}
\end{equation}%
at the Nash equilibrium.

Further more, for the above four cases, we construct the payoff matrix as
obtained from equation (\ref{payoff}) for Alice's four unitary operations as
given in equation (\ref{unitary}).\textbf{\ }For $\gamma =\delta =0$ and $%
\gamma =\delta =\pi /2,$\ the payoff matrix can be obtained from equation (%
\ref{payoff}) as given in table 2. Whereas, for $\gamma =0,$ $\delta =\pi /2$
and $\gamma =\pi /2,$ $\delta =0,$\ the payoff matrix can be obtained from
equation (\ref{payoff}) as given in table 3. We can determine the payoff
which is given to each prisoner on the basis of his strategy from equation (%
\ref{payoff}).

We can see from equations (\ref{unitdef}) and (\ref{unitary}) that each
strategy can be distinguished from the set $U_{A}(\theta _{A},$ $\alpha
_{A}, $ $\beta _{A}),$ $U_{B}(\theta _{B})$ and $U_{C}(\theta _{C}).$ It is
assumed that the two parties Bob and Charlie have a mutual agreement with
each other that they will apply the same strategy in order to find out the
strategy applied by Alice. Let Bob and Charlie apply $\theta _{B(C)}=\pi $,
and gain the payoffs $4,4$ respectively, then they can find that the
decision of Alice was $U_{A}(0,0,0)$ with payoff $0$ as seen from table $2$.
In this way, they can find all the four strategies applied by Alice from
their payoffs, which results an information exchange between the parties
through the arbiter. However, for $\gamma =0,$ $\delta =\pi /2$ and $\gamma
=\pi /2,$ $\delta =0,$\ half of the information is lost because the phase
information vanishes due to the overlapping of half of the entries of the
payoff matrix as seen from table $3$. So there is one half probability to
find out exactly the strategy applied by Alice.

Therefore, from table $2$ and table $3$, we see that Bob and Charlie can
find the unitary operators applied by Alice from their payoffs against their
common strategy. As a result there is a communication of two bits of
information by two local one-qubit operations among the three parties (as
seen from table $2$). However, we can see from table $3$ that the
information shared between the parties is halved because there is one half
probability to find the exact strategy of Alice. Thus, we can establish a
relationship among the amounts of information communicated between the
parties as
\begin{equation}
\{I_{PP}=I_{EE}\}>\{I_{PE}=I_{EP}\}
\end{equation}%
The above relation holds for the set of Alice's four unitary operations
under the bound that Bob and Charlie are restricted to play a common move.

\section{Conclusion}

We present a quantization scheme for three-player Prisoner's Dilemma game
using entangled measuring basis. We show that entanglement plays a dominant
role in a three-player quantum game. We study the communication aspects of a
three-player quantum game which is similar to the dense coding, where, two
bits of classical information can be transmitted by the sender. It is seen
that three-player quantum games are advantageous in the sense that more
information can be carried by the players, thus enhancing the information
flux in comparison to the two-player games. It can be seen that the
communication is due to the advantage of quantum entanglement and quantum
strategies. We investigate that the strategies of the players act as
information carriers in quantum games. We identify four different payoffs on
the basis of different combinations of initial state and measurement basis
entanglement parameters. A relation among these different payoffs is also
established. Exploiting different combinations of initial state and
measurement basis entanglement parameters, we establish a relationship for
the information shared among the parties.\newline

\begin{figure}[tbp]
\begin{center}
\vspace{-2cm} \includegraphics[scale=0.6]{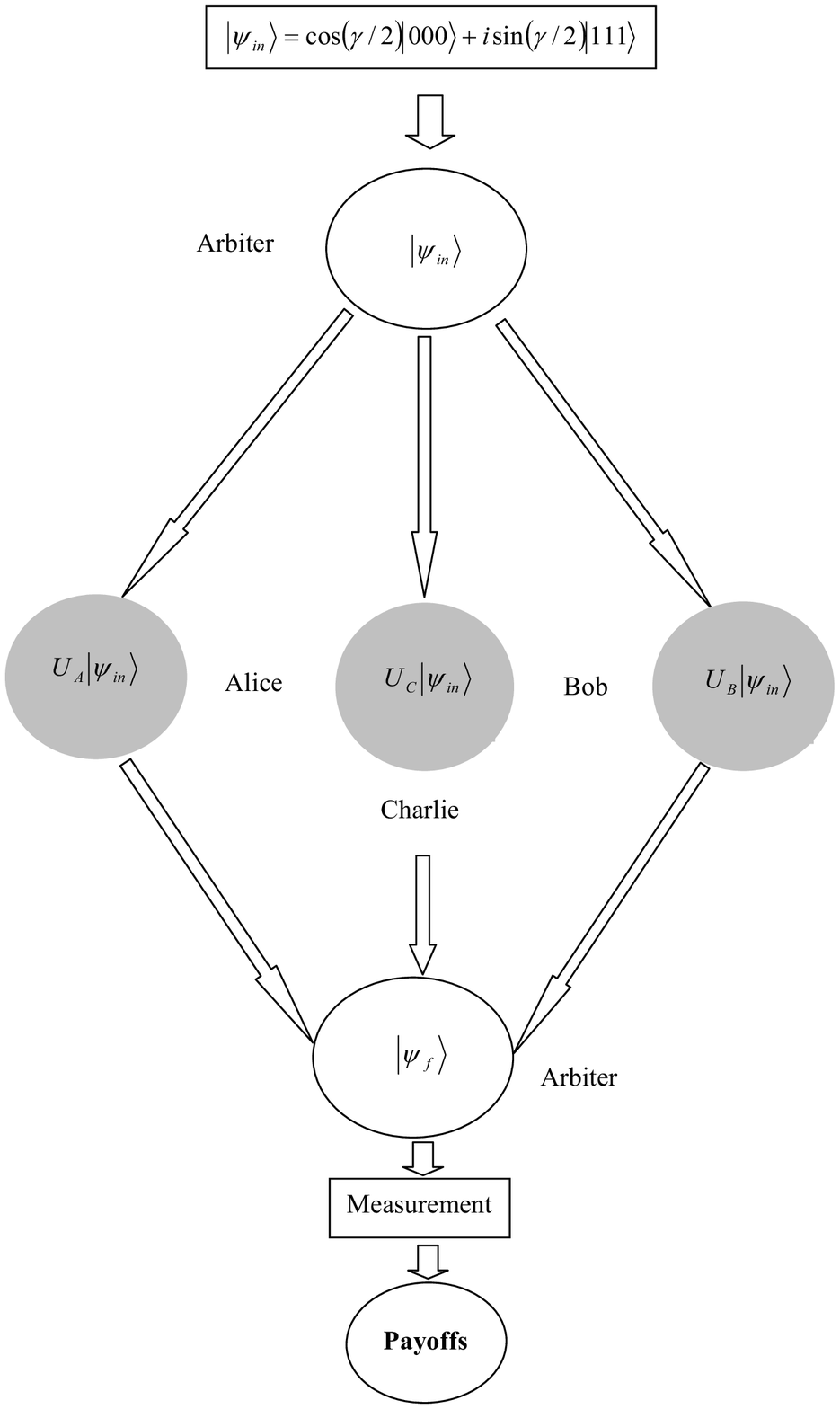} \\[0pt]
\end{center}
\caption{The schematic diagram of the procedure of the game.}
\end{figure}
\newpage

\begin{table}[tbh]
\caption{The payoff matrix for three-player Prisoner's Dilemma where first
number in the parenthesis denotes the payoff of Alice, the second number
denotes the payoff of Bob and the third number denotes the payoff of
Charlie. }$%
\begin{tabular}{c}
\hline
\begin{tabular}{ll}
\ \ \ \ \ \ \ \ \ \ \ \ \ \ \ \ \ \ \ \ \
\begin{tabular}{l}
\underline{Charlie C} \\
\end{tabular}
& \ \ \ \ \ \ \ \ \ \ \ \ \ \ \ \ \ \ \ \ \
\begin{tabular}{l}
\underline{Charlie D} \\
\end{tabular}
\\
\ \ \ \ \ \ \ \ \ \ \ \ \ \ \ \ \ \ \ \ \ \ \ \
\begin{tabular}{l}
\\
\underline{Bob} \\
\end{tabular}
& \ \ \ \ \ \ \ \ \ \ \ \ \ \ \ \ \ \ \ \ \ \ \ \
\begin{tabular}{l}
\\
\underline{Bob} \\
\end{tabular}
\\
$%
\begin{tabular}{l}
\\
\\
\multicolumn{1}{c}{Alice} \\
\\
\end{tabular}%
\begin{tabular}{l|l}
& \ \ \ C \ \ \ \ \ \ \ \qquad D \\ \hline
\begin{tabular}{l}
C \\
\\
D%
\end{tabular}
&
\begin{tabular}{l}
(3,3,3)\qquad (2,5,2) \\
\\
(5,2,2)\qquad (4,4,0)%
\end{tabular}%
\end{tabular}%
$ & $%
\begin{tabular}{l}
\\
\\
\multicolumn{1}{c}{Alice} \\
\\
\end{tabular}%
\begin{tabular}{l|l}
& \ \ \ C \ \ \ \ \ \ \ \qquad D \\ \hline
\begin{tabular}{l}
C \\
\\
D%
\end{tabular}
&
\begin{tabular}{l}
(2,2,5)\qquad (0,4,4) \\
\\
(4,0,4)\qquad (1,1,1)%
\end{tabular}%
\end{tabular}%
$%
\end{tabular}
\\ \hline
\end{tabular}%
$%
\label{di-fit}
\end{table}
\newpage

\begin{table}[tbh]
\caption{The payoffs of the three players for $\protect\gamma =\protect%
\delta =0$ and $\protect\gamma =\protect\delta =\protect\pi /2,$ for
different Alice's operations, as obtained from equation (\protect\ref{payoff}%
).}%
\begin{tabular}{l|l|l|l|l|}
\cline{2-5}
& \multicolumn{2}{|l|}{\ \ \ \ \ \ \ \ \ \ \ \ \ \ \
\begin{tabular}{l}
\underline{$U_{C}(0)$} \\
\end{tabular}%
\ \ \ \ \ \ \ \ \ \ } & \multicolumn{2}{|l|}{\ \ \ \ \ \ \ \ \ \ \ \ \ \ \
\begin{tabular}{l}
\underline{$U_{C}(\pi )$} \\
\end{tabular}%
\ \ \ \ \ \ } \\
& \multicolumn{2}{|l}{} & \multicolumn{2}{|l|}{} \\ \hline
\multicolumn{1}{|l|}{$%
\begin{tabular}{l}
Alice's Unitary \\
Operation%
\end{tabular}%
$} & \ \ \
\begin{tabular}{l}
\underline{$U_{B}(0)$} \\
\end{tabular}
& \ \ \
\begin{tabular}{l}
\underline{$U_{B}(\pi )$} \\
\end{tabular}
& \ \ \
\begin{tabular}{l}
\underline{$U_{B}(0)$} \\
\end{tabular}
& \ \ \
\begin{tabular}{l}
\underline{$U_{B}(\pi )$} \\
\end{tabular}
\\
\multicolumn{1}{|l|}{} &  &  &  &  \\ \hline
\multicolumn{1}{|l|}{\ \
\begin{tabular}{l}
$U_{A}(0,0,0)$ \\
\\
$U_{A}(\pi /3,\pi /2,\pi /2)$ \\
\\
$U_{A}(\pi /2,\pi /2,\pi /2)$ \\
\\
$U_{A}(\pi ,\pi ,\pi )$%
\end{tabular}%
} & \ \ \
\begin{tabular}{l}
(3,3,3) \\
\\
(3/4,7/4,7/4) \\
\\
(1/2,5/2,5/2) \\
\\
(5,2,2)%
\end{tabular}
& \ \ \ \
\begin{tabular}{l}
(2,5,2) \\
\\
(7/2,1/2,17/4) \\
\\
(3,1,9/2) \\
\\
(4,4,0)%
\end{tabular}
& \ \ \ \
\begin{tabular}{l}
(2,2,5) \\
\\
(7/2,17/4,1/2) \\
\\
(3,9/2,1) \\
\\
(4,0,4)%
\end{tabular}
& \ \ \ \
\begin{tabular}{l}
(0,4,4) \\
\\
(9/2,9/4,9/4) \\
\\
(4,5/2,5/2) \\
\\
(1,1,1)%
\end{tabular}
\\
\multicolumn{1}{|l|}{} &  &  &  &  \\ \hline
\end{tabular}%
\end{table}
\newpage

\begin{table}[tbh]
\caption{The payoffs of the three players for $\protect\gamma =0,$ $\protect%
\delta =\protect\pi /2$ and $\protect\gamma =\protect\pi /2,$ $\protect%
\delta =0,$ for different Alice's operations, as obtained from equation (%
\protect\ref{payoff}).}%
\begin{tabular}{l|l|l|l|l|}
\cline{2-5}
& \multicolumn{2}{|l|}{\ \ \ \ \ \ \ \ \ \ \ \ \ \ \
\begin{tabular}{l}
\underline{$U_{C}(0)$} \\
\end{tabular}%
\ \ \ \ \ \ \ \ \ \ } & \multicolumn{2}{|l|}{\ \ \ \ \ \ \ \ \ \ \ \ \ \ \
\begin{tabular}{l}
\underline{$U_{C}(\pi )$} \\
\end{tabular}%
\ \ \ \ \ \ } \\
& \multicolumn{2}{|l}{} & \multicolumn{2}{|l|}{} \\ \hline
\multicolumn{1}{|l|}{$%
\begin{tabular}{l}
Alice's Unitary \\
Operation%
\end{tabular}%
$} & \ \ \
\begin{tabular}{l}
\underline{$U_{B}(0)$} \\
\end{tabular}
& \ \ \
\begin{tabular}{l}
\underline{$U_{B}(\pi )$} \\
\end{tabular}
& \ \ \
\begin{tabular}{l}
\underline{$U_{B}(0)$} \\
\end{tabular}
& \ \ \
\begin{tabular}{l}
\underline{$U_{B}(\pi )$} \\
\end{tabular}
\\
\multicolumn{1}{|l|}{} &  &  &  &  \\ \hline
\multicolumn{1}{|l|}{\ \
\begin{tabular}{l}
$U_{A}(0,0,0)$ \\
\\
$U_{A}(\pi /3,\pi /2,\pi /2)$ \\
\\
$U_{A}(\pi /2,\pi /2,\pi /2)$ \\
\\
$U_{A}(\pi ,\pi ,\pi )$%
\end{tabular}%
} & \ \ \
\begin{tabular}{l}
(2,2,2) \\
\\
(17/8,9/4,9/4) \\
\\
(9/4,5/2,5/2) \\
\\
(5/2,3,3)%
\end{tabular}
& \ \ \ \
\begin{tabular}{l}
(3,5/2,3) \\
\\
(3,21/8,23/8) \\
\\
(3,11/4,11/4) \\
\\
(3,3,5/2)%
\end{tabular}
& \ \ \ \
\begin{tabular}{l}
(3,3,5/2) \\
\\
(3,23/8,21/8) \\
\\
(3,11/4,11/4) \\
\\
(3,5/2,3)%
\end{tabular}
& \ \ \ \
\begin{tabular}{l}
(5/2,3,3) \\
\\
(19/8,11/4,11/4) \\
\\
(9/4,5/2,5/2) \\
\\
(2,2,2)%
\end{tabular}
\\
\multicolumn{1}{|l|}{} &  &  &  &  \\ \hline
\end{tabular}%
\end{table}


\begin{thebibliography}{}
\bibitem{Nelson} Nielson M A and Chuang I L 2000 Quantum Computation and
Quantum Information (Cambridge: Cambridge University Press)

\bibitem{Dirk} Bouwmeester D, Ekert A and Zeilinger A 2000 The Physics of
Quantum Information (Berlin: Springer Verlag)

\bibitem{Meyer} Meyer D\ A 1999 Phys. Rev. Lett. \textbf{82} 1052

\bibitem{Flitney} Flitney A P and Abbott D 2002 Fluct. Noise Lett. \textbf{2}
R175

\bibitem{Lee} Lee C F and Johnson N 2003 Phys. Rev. A \textbf{67} 022311

\bibitem{Du} Du J et al 2003 J. Phys. A: Math. Gen. \textbf{36} 6551

\bibitem{Nawaz} Nawaz A and Toor A\ H 2006 J. Phys. A: Math. Gen. \textbf{39}
2791

\bibitem{Lee2} Lee C\ F, Johnson N\ F 2003 Physics Letters A \textbf{319} 429

\bibitem{Eisert} Eisert J, Wilkens M, Lewenstein M 1999 Phys. Rev. Lett.
\textbf{83} 3077

\bibitem{Du1} Du J, Li H, Xu X, Shi M, Wu J, Zhou X and Han R 2002 Phys.
Rev. Lett. \textbf{88} 137902

\bibitem{Prev} Prevedel R, Stefanov A, Walther P and Zeilinger A 2007 New J.
Phys. \textbf{9} 205

\bibitem{Benj} Benjamin S\ C and Hayden P\ M 2001 Phys. Rev. A \textbf{64}
030301(R)

\bibitem{Du2} Du J et al. 2002 Phys. Lett. A \textbf{302} 229

\bibitem{Flitney2} Flitney A\ P and Abbott D 2004 J. Opt. B: Quantum
Semiclass. Opt. \textbf{6} S860

\bibitem{Li} Li H, Du J F and Massar S 2002 Phys. Lett. A \textbf{306} 73

\bibitem{Ramzan} Ramzan M, Nawaz A, Toor A\ H\ and Khan M K 2008 J. Phys.\
A: Math. Theor. \textbf{41} 055307

\bibitem{Kawakami} Kawakami T 2002 Lecture Notes in Computer Science
(Berlin: Springer)

\bibitem{Orlin} Grabbe J O 2005 Preprint: quant-ph/0506219

\bibitem{King} King C and Ruskai M B 2001 J. Math. Phys. \textbf{42} 87

\bibitem{Nawaz2} Nawaz A and Toor A\ H 2006 J. Phys. A: Math. Gen \textbf{37}
11457

\bibitem{Coding} Bennett C\ H and Wiesner S\ J 1992 Phys. Rev. Lett. \textbf{%
69} 2881.

\bibitem{Patil} Pati A K and Agrawal P 2004 J. Opt. B: Quantum Semiclass.
Opt. \textbf{6} S844

\bibitem{Kim} Kim Y H, Kulik S P and Shih Y 2001 Phys. Rev. Lett. \textbf{86}
1370\pagebreak \newline
{\LARGE Figure Caption}\newline
\textbf{Figure 1}. The schematic diagram of the procedure of the game.%
\newline
{\LARGE Tables Captions}\newline
\textbf{Table 1}. The payoff matrix for three-player Prisoner's Dilemma
where first number in the parenthesis denotes the payoff of Alice, the
second number denotes the payoff of Bob and the third number denotes the
payoff of Charlie.\newline
\textbf{Table 2}. The payoffs of the three players for $\gamma =\delta =0$
and $\gamma =\delta =\pi /2,$ for different Alice's operations, as obtained
from equation (\ref{payoff}).\newline
\textbf{Table 3}. The payoffs of the three players for $\gamma =0,$ $\delta
=\pi /2$ and $\gamma =\pi /2,$ $\delta =0,$ for different Alice's
operations, as obtained from equation (\ref{payoff}).\newpage
\end{thebibliography}
\end{document}